\documentstyle[prl,aps,floats,psfig]{revtex}

\ifpreprintsty\else\textheight = 57\baselineskip\fi

\def\DDD{\delta}

\def\krp{K_{\ch,+}}
\def\kr{K_{\ch}}

\def\tdis{T_{dis}}

\def\three#1{}

\def\srcax{Sr$_{14-x}$Ca$_{x}$Cu$_{24}$O$_{42-\delta}$ }
\def\srca{Sr$_{0.4}$Ca$_{13.6}$Cu$_{24}$O$_{42-\delta}$}

\newcommand{\cmin}{\lesssim} 
\newcommand{\cmag}{\gtrsim}

\def\ch{\rho}
\def\sp{\sigma}
\def\gsig{s}

\def\potpert{perturbation }
\def\potperts{perturbations }

\def\nc{n_c}

\def\beq{\begin{equation}}
\def\eeq{\end{equation}}
\def\beqn{\begin{eqnarray}}
\def\eeqn{\end{eqnarray}}

\def\eqref#1{ Eq. (\ref{#1})}

\newcommand{\ferd}{}
\newcommand{\fer}{}
\newcommand{\bosp}{}
\newcommand{\bost}{}
\newcommand{\nnpt}{}

\newcommand{\ph}{\varphi}
\def\th{\theta}

\newcommand{\setferq}{
\renewcommand{\ferd}[2]{\psi^{\dag}_{##1}(##2)}
\renewcommand{\fer}[2]{\psi_{##1}(##2)}
}

\setferq %

\newcommand{\lbosp}[1]{\ph^{(#1)}}
\newcommand{\lbost}[1]{\th^{(#1)}}

\newcommand{\nnb}[2]{n^{#1}_{#2}}

\newcommand{\setbostr}{
\renewcommand{\bosp}[1]{\ph_{##1}}
\renewcommand{\bost}[1]{\th_{##1}}
\renewcommand{\nnpt}[2]{\nnb{##1}{##2}}
}

\setbostr %

\def\QQ{\nu} 
\def\PP{p}

\long\def\bibdef#1#2#3{\newenvironment{#1}{#2}{}}
\def\lcit#1{\begin{#1}\end{#1}}
\def\nonformale#1{#1}
\def\formale#1{}
\def\spa{\vspace*{-.5cm}} \def\spb{\vspace*{-1.6cm}}

\bibdef{no.wh.95}{
R.~M. Noack, S.~R. White, and D.~J. Scalapino, Europhys. Lett. {\bf 30},  163
  (1995)}%

\bibdef{scal.nature}{
D.~J. Scalapino, Nature {\bf 377},  12  (1995)}%

\bibdef{ec.ba.94}{
R.~S. Eccleston {\it et~al.}, Phys. Rev. Lett. {\bf 73},  2626  (1994)}%

\bibdef{ko.ta.95}{
K. Kojima {\it et~al.}, Phys. Rev. Lett. {\bf 74},  2812  (1995)}%

\bibdef{ka.fi.92}{
C.~L. Kane and M.~P.~A. Fisher, Phys. Rev. B {\bf 46},  15233  (1992)}%

\bibdef{fu.na.93}{
A. Furusaki and N. Nagaosa, Phys. Rev. B {\bf 47},  4631  (1993)}%

\bibdef{emer.rev}{
V.~J. Emery,  in {\em Highly-Conducting One-Dimensional Solids}, edited by
  J.~T. Devreese, R.~E. Evrard, and V. van Doren (Plenum Press, New York,
  1979)}%

\bibdef{hald.81}{
F.~D.~M. Haldane, J. Phys. C {\bf 14},  2585  (1981)}%

\bibdef{or.gi.96}{
E. Orignac and T. Giamarchi, Phys. Rev. B {\bf 53},  R10453  (1996)}%

\bibdef{kh.ri.94}{
D.~V. Khveshchenko and T.~M. Rice, Phys. Rev. B {\bf 50},  252  (1994)}%

\bibdef{fabr.93}{
M. Fabrizio, Phys. Rev. B {\bf 48},  15838  (1993)}%

\bibdef{schu.96}{
H.~J. Schulz, Phys. Rev. B {\bf 53},  R2959  (1996)}%

\long\def\singlecol#1{
\twocolumn[\hsize\textwidth\columnwidth\hsize\csname @twocolumnfalse\endcsname
              #1]}

\ifpreprintsty
\long\def\singlecol#1{#1}
\fi

\long\def\beginfig#1#2{\begin{figure}[htb] \centerline{#1}
    \protect{#2}
 \end{figure}}
\def\endfigures{}

\ifpreprintsty
      \long\def\beginfig#1#2{}
      \def\endfigures{
           \long\def\beginfig##1##2{\begin{figure}  ##2 \end{figure}}
\figa 
\figb
}
\fi

\newcommand{\pcite}[1]{\cite{#1}}

\begin{document}

\newcommand{\figa}{
\beginfig{
\psfig{width=8.3truecm,file={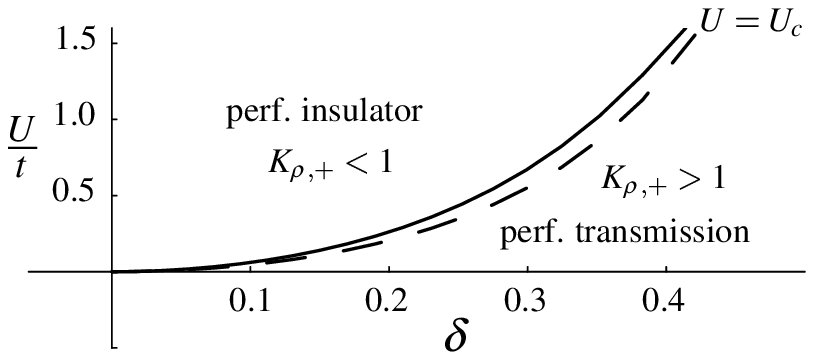}}}{\caption{ 
 Phase diagram for the two-chain system with a
  barrier or with a low  impurity concentration.
The two estimates for the 
boundary $U=U_c$
are obtained
according to contribution (i) only (dashed) and additionally (ii) (full) as explained in
the text.
\label{fig1}
}}}

\newcommand{\figb}{
\beginfig{
\psfig{width=8.3truecm,file={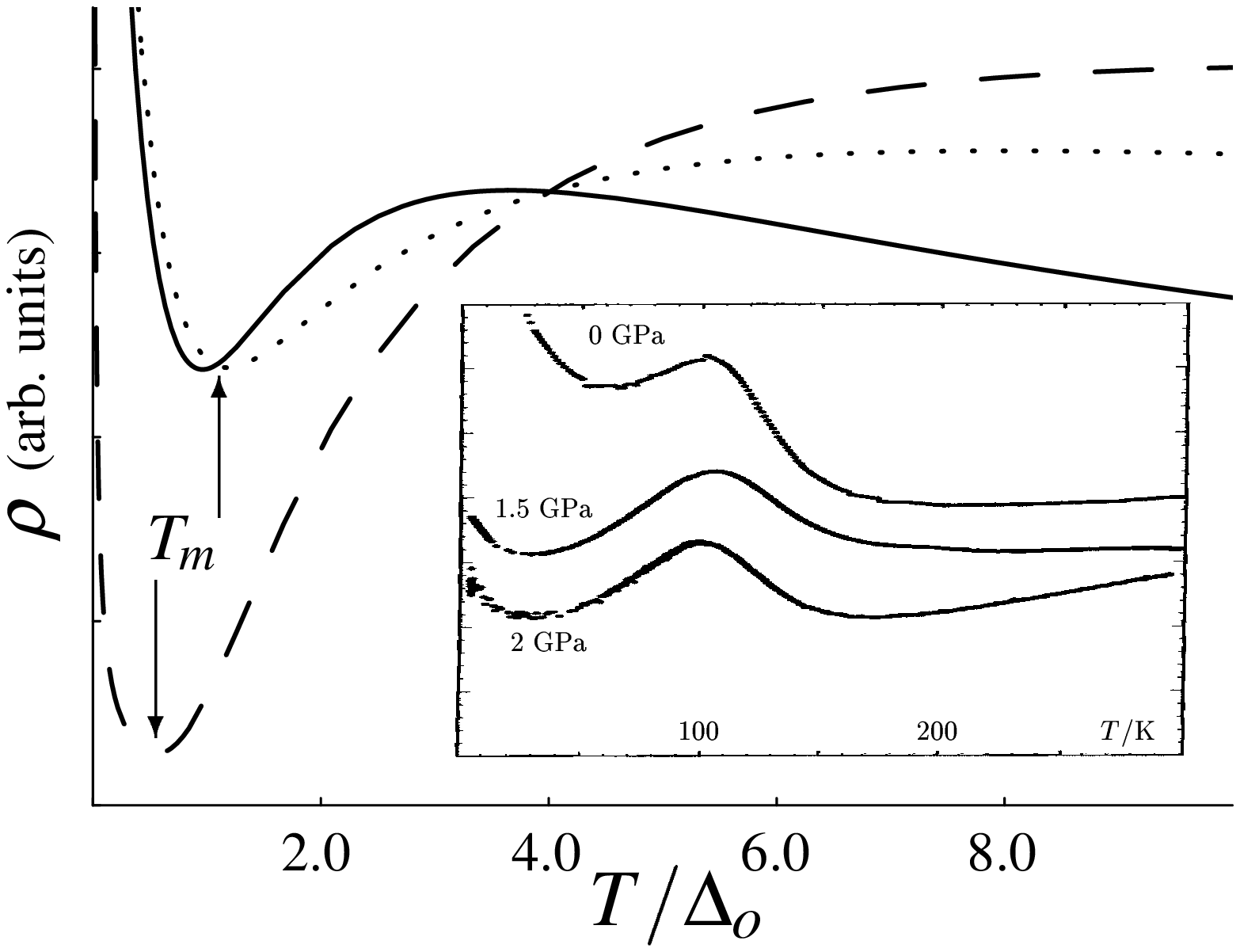}}
}
{
\caption{
Behavior of the resistivity $\rho$ as a function of temperature 
(in units of the odd-mode gap $\Delta_o$)
for  values of the parameters (as explained in the text):
 $\tilde v=0.4$ and $\Delta_s/\Delta_o=4.$
  (full line);  $\tilde v=0.4$ and $\Delta_s=\Delta_o$
(dotted);
$\tilde v=0.1$ and $\Delta_s=\Delta_o$ (dashed). 
For all curves, $\kr=\krp=0.6$ and $E_F/\Delta_o=150$.
Inset: $\rho(T)$ for the normal state of 
\srcax from Ref. [8] for different values of
pressure.
(With kind permission of the authors).
}
\label{fig2}
}
}

\draft   %

\title{ 
Electron transport in coupled chains of interacting fermions with
impurities
}

\author{E. Arrigoni, B. Brendel and W. Hanke 
}
\address{ 
Institut f\"ur Theoretische Physik,
Universit\"at W\"urzburg,
D-97074 W\"urzburg, Germany
}

\singlecol{

\maketitle

\begin{abstract}
We study the low-temperature transport of a doped two-chain
ladder system of interacting fermions in the presence of a barrier or
of a low concentration of impurities. Above a certain value of the
interaction, the conductance is suppressed, like for a single chain,
despite the presence of dominant superconducting correlations. There
is, however, a region of repulsive interaction where perfect
transmission across the barrier occurs unlike the
single-chain case.
We provide a possible explanation for
the temperature maximum of the
resistivity in the normal state of \srca.
\end{abstract}


}%

Transport properties in one-dimensional (1D) interacting fermion
systems, i.e. Luttinger liquids
\pcite{hald.81},
are remarkably affected by  
the interaction.
 Indeed, it has been shown\cite{ka+fu}
  that the linear
conductance across an  impurity is {\it completely suppressed}  in the
zero-temperature limit for repulsive  interaction
{\it no matter how weak the barrier potential}.
Recently, the question of what happens when a small number of
Luttinger Liquids are coupled
by a single-particle hopping
 has received particular 
interest\pcite{ri.go.93,no.wh+scal,twoc,ba.fi.96}.
Coupled chains
 represent the first step towards a two-dimensional
system, but  the important effects of one-dimensionality in the single
chains can still be  treated in an accurate
way. 
Various numerical
 calculations and renormalization-group (RG) treatments
 agree about the fact that the two-chain system 
is in
a spin-gapped state with dominant ``odd-mode'' (the analogy of
 d-wave in two chains) 
superconducting 
correlations\pcite{ri.go.93,no.wh+scal,twoc,ba.fi.96}.
The apparent 
similarity with the high-T$_c$ superconductors has suggested the idea 
that  the pairing mechanism 
in the high-T$_c$ materials 
could be of a similar origin.
This theoretical interest
is supplemented by
experimental realizations of systems of 2 to 5 coupled chains, such
as VO$_2$P$_2$O$_7$, 
Sr$_{n-1}$Cu$_{n+1}$O$_{2n}$, and 
others\pcite{ec.ba+ko.ta}.
In addition to the observation of a remnant spin gap in the doped
metallic state,
superconductivity has been recently observed at high pressure
in the ladder material 
\srca\pcite{ue.na.96}.
Finally, narrow GaAs quantum wires with few transverse
channels can also be represented by coupled-chain systems 
in the intermediate
situation between one and many channels\pcite{timp.90}.

In this Letter, we 
 study  how the results of Refs. \onlinecite{ka+fu} 
for the transport of interacting electrons across an impurity
in a 1D Luttinger liquid
evolve
 when two 1D systems are coupled together.
In particular, we  analyze
 the 
influence 
of the  gaps (in the spin and in the ``odd'' charge modes) 
 and of the  superconducting
correlations,
present in the two-chain system,
on  transport properties.
Specifically, we 
study the conductance of a system of two 
Hubbard-like chains coupled by a single-particle hopping 
in the presence of a
barrier or a low impurity concentration.
This system is a good
description for a two-channel conductor and 
for the  ladder materials.
Similarly to the pure 1D
case\pcite{ka+fu},
 we obtain that the conductance across the impurity is completely
 suppressed whenever the ``even charge'' 
correlation exponent $\krp$ is less than unity\pcite{notation}. 
However,  there is an additional 
  region of {\it repulsive}
electron-electron interaction, 
$U \cmin U_c$, 
where 
 $\krp>1$ and 
perfect transmission
 across the impurity occurs.
A similar result is obtained for the three-chain case where,
  in the pure case,
more gapless modes are 
present
\pcite{3c.pla,unpub}.
If we limit ourselves to weak impurities and temperature $T$ smaller
 than all the gaps, then our results for the $T$-dependence of the
 conductivity $\sigma$ agree with a recent treatment of a large
 concentration of weak impurities \onlinecite{or.gi}.

We consider a Hamiltonian of the general form
\beq
H= H_{chain,t} + H_{chain,U}  + H_{\perp} + V_b \;,
\eeq
where 
$ H_{chain,t}+H_{chain,U} $ describes $n_c$
($=2$) uncoupled 1D Hubbard chains, with
 $H_{chain,t}$ 
denoting the intra-chain hopping $t$, and 
$ H_{chain,U}$ the
 on-site repulsion $U$. 
The chains are coupled by an inter-chain
 hopping term $ H_{\perp}$ with hopping amplitude
$t_{\perp}$ 
(with open boundary conditions in the perpendicular   %
direction).
 $V_b$ is the impurity potential described in detail below.
It  should be pointed out that the qualitative results  do not depend
significantly  
on the detailed structure of the model. 
One can easily extend the method and the results to other
 systems  with, e. g., a more general band structure than the simple
 cosine band and a 
nearest-neighbor interchain 
and/or intrachain interaction.    %
In the weak-coupling case, one first diagonalizes the single-particle
part of the Hamiltonian ($ H_{chain,t} + H_{\perp}$),
obtaining $\nc$ bands. The interaction $H_{chain,U}$ written in the
band representation includes different classes of 
 processes that scatter  particles or holes at the ``Fermi surface''
(which consists of $2\nc$ points). 
These interactions have to be treated by a RG
analysis, since perturbation theory is divergent. The clean case
($V_b=0$)
 has been
considered
 by several authors (Refs. \onlinecite{twoc,ba.fi.96} for 
two chains  
and 
\onlinecite{3c.pla}
for  three chains).

One introduces
four boson fields  $\lbosp{\gsig,q}$ and four dual fields 
$\lbost{\gsig,q}$
for the two spins $\gsig=\pm 1$ and for the two bands
$q=0,\pi$. Taking
  the symmetric and antisymmetric combinations of the band and of
the spin part, one obtains the fields 
$\bosp{\ch,\pm}$, $\bosp{\sp,\pm}$, $\bost{\ch,\pm}$, and
$\bost{\sp,\pm}$\pcite{notation}.  
The RG calculations of Refs.
\onlinecite{twoc,ba.fi.96}  yield a mass gap with
short-range correlations for both ``spin'' 
degrees of freedom $\bosp{\sp,\pm}$ and for the 
 ``odd-charge'' degree of freedom $\bost{\ch,-}$\pcite{c1s0}. 
In this phase, ``odd mode'' pairing correlations 
decay rather slowly (as $r^{-1/(2\krp)}$) and  dominate with
respect to other correlations in the weak-coupling limit.

Next, we introduce the \potpert due to an impurity in a single chain, or due to a
symmetric ``barrier''  in both chains:
\beq
V_b = \int \ dx \ \sum_{q_1,q_2,\gsig} V_{q_1,q_2}(x)
\ferd{\gsig,q_1}x  \fer{\gsig,q_2}x \;,
\label{vb}
\eeq
where $\fer {\gsig,q}x$ destroys a fermion with spin $\gsig$ on band
$q$ 
at site $x$. 
Here, the impurity potentials $V_{q_1,q_2}(x)$ satisfy
 $V_{0,0}(x)=V_{\pi,\pi}(x)$ and, for a symmetric barrier, 
 $V_{0,\pi}(x)=0$.
Introducing the boson representation and
considering a potential that is
restricted to a small region around $x=0$,
 one obtains three
contributions for the  \potpert: a backscattering intraband term, i.e.\ 
\beq
V_{intra} \sim 
\sum_{\gsig} v_{0,0}(2 k_F) \ 
\cos(\bosp{\ch,+}+\gsig\bosp{\sp,+}) 
  \cos(\bosp{\ch,-}+ \gsig \bosp{\sp,-}) 
\label{vintra}
\eeq
and two interband  (back- and forward-scattering) terms
(the forward-scattering intraband term can, as usual,
be absorbed by a canonical transformation).
The interband terms 
 have a form
similar to $V_{intra}$, except that 
 some cosines are replaced with sines and
the phase fields
$\bosp{\QQ,-}$ (for the backscattering) or $\bosp{\QQ,+}$
(for the forward scattering) are replaced with
$\bost{\QQ,-}$ ($\QQ=\ch$ or $\sp$).
In \eqref{vintra},
  $v_{0,0}(k)$ is proportional 
 to the Fourier transform 
of $V_{0,0}(x)$
 with longitudinal momentum $k$
 [we have approximated 
$v_{0,0}(2k_F)$ by $v_{q,q}(2 k_F(q))$], and 
 the fields are
implicitly considered to be at the origin, $x=0$.

As pointed out in Ref.\onlinecite{ka+fu}, the  \potpert will
evolve under RG
 into a more general form
consistent with the symmetries of the Hamiltonian.
In our case, $V_b$ is symmetric under rotation by $2 \pi$ of
any of the fields $\bosp{\QQ,+}$, $\bosp{\QQ,-}$, or
$\bost{\QQ,-}$.
Moreover,  a simultaneous rotation of
$\pi/2$ of {\it all} six of these fields or a simultaneous
rotations by $\pi$ of some groups of two or three fields (such as, e. g.,
$\bosp{\ch,+}$, $\bosp{\ch,-}$, and $\bost{\sp,-}$) leaves the
 \potpert (and also the rest of the Hamiltonian) invariant. 
The general \potpert operator which is consistent with these symmetries 
develops under the RG and can be cast into the form
\beq
\label{genpert}
V_b' = %
 \sum_{\nnpt{\ph}{\QQ,\pm},\nnpt{\th}{\QQ,-}}
 {\cal V}(\underline{n})
 \exp\left[i  \sum_{\QQ=(\ch,\sp)} \left( \nnpt{\th}{\QQ,-}\bost{\QQ,-} +
               \sum_{\PP=\pm}
               \nnpt{\ph}{\QQ,\PP}\bosp{\QQ,\PP}\right)\right] \;.
\eeq
Here, the ``Fourier coefficients'' ${\cal V}(\underline{n})$
are nonzero only for values of the integers
$\underline{n}\equiv \{\nnpt{\{\ph,\th\}}{\QQ,\pm}\}$
 satisfying certain constraints imposed by
symmetries. 
A calculation, in principle similar to ours,
has been carried out by T. Kimura and coworkers 
for the two-chain case  \cite{ki.ao.95}.
However, these authors restrict their analysis to 
 the {\it bare} 
\potperts [Eq. (\ref{vintra}) plus  the interband terms],
which are completely suppressed  due to the presence of 
the gap in the $\bost{\ch,-}$ and $\bosp{\sp,-}$ field (at least
in  phases I and   %
III of Ref. \onlinecite{ki.ao.95}).
 Thus, they 
argue that the system should behave as a perfect conductor even in the
presence of impurities.
We show here, however, that their result is invalid if one carefully
takes into account the additional terms in \eqref{genpert} generated
               under the RG. 
Indeed, perturbation terms where the subset of integers 
$\nnpt{\ph}{\ch,-}$ and $\nnpt{\th}{\sp,-}$
vanishes survive.
 The most relevant
terms are the ones with the
smallest $\nnpt{\ph}{\ch,+}$ consistent with the constraints on the
$\nnpt{\{\ph,\th\}}{\QQ,\pm}$ discussed above.
One (nontrivial) solution 
is given, for
example,
by the \potpert term  %
$
v_2 \cos(2 \bosp{\ch,+}) \;.
$
Here, 
we have exploited  inversion symmetry
and replaced the gapped fields with their average (vanishing) values.
To see how such a term can originate,  consider the 
intraband part of
the initial \potpert in \eqref{vintra}. 
Although its coefficient $v_0(E)$ (whose {\it bare} value
is  $v_0(E_F)\equiv v_{0,0}(2 k_F)$)
is suppressed during the RG transformation\pcite{ki.ao.95}, 
$v_2$
 has the  RG equation
\beq
\label{rgv2}
\frac{d\ v_2(E) }{-d\ \log E} 
  = -\frac18 
[v_0(E)]^2 + 
(1- \krp) v_2(E) \;,
\eeq
where $E$ is the characteristic energy parameterizing the RG 
flow.
Therefore $v_2$, although having vanishing initial conditions $v_2(E=E_F)=0$,
is  obtained under RG from the irrelevant term
$ v_0(E)$ 
and is itself {\it relevant} whenever  $\krp<1$.
Physically, the fermionic representation of the
operator associated with $v_2$ is  \mbox{$\quad \ferd{+,\gsig,0}x
\ferd{+,\gsig,\pi}x 
\fer{-,\gsig,\pi}x
\fer{-,\gsig,0}x$}\pcite{notation}, 
which represents an {\it interband electron-pair
backscattering operator}. This term is absent in the bare impurity
potential,   but is obtained  to
second-order perturbation theory in $V_{intra}$.

When $\krp>1$, the perturbation 
$v_2$
becomes irrelevant and the
low-temperature ($T$ smaller than all the gaps)  %
conductance $G(T)$ behaves as
$\DDD G(T)\equiv G(T)-G_0 \propto - v_2(T)^2 \propto - v_0^4
  T^{2(\krp-1)}$ \pcite{v0v2}
where $G_0$ is the value
of the conductance without impurities, $v_0\equiv v_0(E_F)$,
 and $v_2(T)$ is the solution of \eqref{rgv2}.
For $\krp<1$, the expression for $\delta G(T)$
is valid only down to a temperature $T=T^*$
 with 
$\DDD G(T^*)/G_0\cmin 1$.
For a low  impurity concentration $n$, such that the temperature is
larger than the ``discretization energy'' $\tdis\equiv v_F n$, 
the  impurities can effectively be treated independently and their
contribution to the total resistance adds up incoherently\pcite{ka+fu}.
Thus, the
resistivity $\rho$ behaves like
$\rho \propto v_0^4\ n\ T^{2(\krp-1)}$
(down to  $T=\max(T^*,T_{dis})$), 
 in agreement with  
 Ref.\onlinecite{or.gi}.

When $\krp<1$ and $T<T^*$, 
the renormalized \potpert diverges
and one has to
treat the case of  large 
$v_2$. 
In this limit, and for a single impurity, we can  fermionize back
the operator associated with 
$v_2$
and consider two decoupled 
semi-infinite
(spinless)  Luttinger Liquids
connected via a weak tunneling matrix element
$t_2$\pcite{ka+fu}.
This gives 
a perturbation dual to $v_2$
:
$t_2 \cos(2 \bost{\ch,+}(0))$.
Similarly to \cite{ka+fu}, 
one obtains a
 conductance 
$G(T) \propto  t_2^2 \  T^{(2/\krp)-2}$, 
which vanishes at low temperatures.

Therefore, the transmission through an impurity in a  two-chain
system is similar to the case of a single chain. 
For 
$\krp<1$
the impurity is  perfectly insulating, whereas perfect transmission
occurs 
in the case of 
$\krp>1$,  {\it independent of the barrier's strength and 
 symmetry}.
Physically, this can be understood by considering 
that 
$v_2$
is associated with
a charge-density wave (CDW) operator
with wave vector \mbox{$2 k_{F}(0)+2
k_F(\pi)\equiv 4 k_F$} whose correlations
 decay like $r^{-2 \krp}$.
The argument of Ref.\onlinecite{ka+fu} can thus be applied to these 
$4 k_F$ CDW
correlations (since the ``regular'' 
$2k_F$ ones are suppressed\cite{ki.ao.95}).
 When  $\krp<1$ 
 the   $4 k_F$ CDW correlations 
 are long-range enough and are pinned by
 an arbitrarily weak impurity at low energies.

Nevertheless,
there is an essential
 difference between the single- and the two-chain system, since
$\krp$ is renormalized 
by an amount
$\DDD \krp \propto (\DDD v_F/v_F)^2$ 
by the
presence of the gap $\Delta_o$ in the ``odd'' modes \cite{ba.fi.96} %
(here, $\DDD v_F$ is  the difference  between the
 Fermi velocities $v_F$ of the bonding and antibonding bands).
This renormalization can be considered as an effective 
attraction mediated by the odd gapped modes.
The RG equation for $\krp$ as a function of the flux parameter   %
$\Gamma$ ($\Gamma$ is proportional to $l=-\log E/E_F$ in our notation)
is reported in Ref. \onlinecite{ba.fi.96} 
(in order to match our notation,  one has to replace $\krp \to
1/\krp$ in that reference).
The exact value of $\DDD \krp$ cannot be determined 
even in the $U/t\to 0$ limit, since it depends on the  point 
$l^*$ where
 the RG flow is stopped.
A reasonable lower bound and 
estimate for $\DDD \krp$ and for the critical value
$U_c$, which separates regions $\krp<1$ and $\krp>1$ in
Fig. 1 can be obtained in the following way:
Notice that $\DDD \krp$ contains
two contributions:
(i) an $l$-independent contribution 
 obtained by neglecting the RG flow of the couplings
(i. e. by taking $l=0$ and by integrating out the gapped modes), and
(ii) an additional contribution
 due to the renormalization of the couplings which
depends on $l$.
One can show that (i) sets a {\it lower} bound for 
$\DDD \krp$ and that the contribution of (ii) is much smaller
than (i) and only {\it weakly} $l$-dependent.

In Fig. 1, we plot  $U=U_c$
as a function of doping $\delta$ (away from half-filling) for the isotropic
case $t_{\perp}=1$, by taking into account in the dashed line 
only the contribution
(i)  and by including in the full line additionally 
the contribution (ii)\cite{uc}.
 For large doping there is a sizable region
 of repulsive interaction for which the transmission across the impurity
 is perfect. This effect will be  particularly relevant for
 systems in which the Hubbard repulsion $U\cmin t$, as for example
 in carbon nanotubes\cite{ba.fi.cn.96} and in GaAs quantum wires.
Exact numerical results for small $U$ \cite{ya.sh.95}  
yield the  %
opposite effect, namely $\krp$ seems to be reduced with respect to the
single-chain value. 
We believe, however, that  in
these numerical studies
the value of $\krp$ is underestimated
since the system size considered ($6$ lattice spacings $a_0$)
is smaller than the spin correlation
length $\xi_s \propto v_F/\Delta_s$, which gives $\xi_s \cmag 30-40
a_0$\pcite{no.wh+scal}, where $\Delta_s$ is the spin gap.

\figa

For a finite impurity concentration,
 the resistivity $\rho(T)$
  diverges at low 
temperatures in the region in which $\krp<1$, whereas  it  
decreases
in the region  in which $\krp>1$ \pcite{tdis}.
At temperatures larger than the gap, $T\gg 
\Delta_o \approx \Delta_s$ \pcite{ident},
the resistivity is dominated by the scattering
 with the $2k_F$ potential $v_0$. 
Its behavior\cite{valid}
is
$\rho(T) \propto v_0^2 n T^{(\kr -1)/2}$, 
 yielding  an increasing 
 resistivity with decreasing $T$ (for $U>0$). 

For a more quantitative evaluation of $\rho(T)$
and in order to find the behavior of   $\rho(T)$
 in the crossover regime for %
$T$ of the order of  $\Delta_o$, we use
the perturbative expression
$\rho(T)\propto c_0 \ v_0(T)^2 +
 c_2 \ v_2(T)^2$, where $c_0$ and $c_2$ are constants \cite{c2c0}.
For these temperatures, $v_0(T)$ can be described by an   
  activated behavior $v_0(T) \approx  (T/E_F)^{(\kr -1)/4}  v_0
  \exp(\Delta_o/E_F-\Delta_o/T)$, whereas
$v_2$ can be evaluated from \eqref{rgv2} \pcite{unpub}.
In the weak-impurity regime which we are considering, the parameter
$\tilde v\equiv v_0 \sqrt{c_2/c_0}$\cite{c2c0}
takes a value $\tilde v\cmin 1$. In this case
$\rho(T)$ is dominated by the $2k_F$-scattering potential $v_0$
from higher temperatures 
down to  a crossover temperature
$T=T_{m}\cmin\Delta_o$, with  $T_m \sim \Delta_o/(2\log(2/\tilde v))$.
Therefore, $\rho(T)$ 
eventually decreases
 down to
 this crossover temperature
$T_{m}$.
For $T<T_m$  the
$4k_F$-scattering $v_2$ becomes dominant
and, if $\krp<1$,
$\rho(T)$ dramatically
 increases  eventually
diverging for low $T$.

In Fig. 2,  we plot the behavior of $\rho(T)$
for different values
 of the parameters 
$\tilde v$ and $\Delta_s/\Delta_o$, and
for
$\kr\approx\krp=0.6$, which
 corresponds to $U/t\approx 8.0$\cite{params} and doping $\delta\sim 
0.25$\pcite{tdis,os.mo.96}.
The full line is obtained by taking a large ratio
$\Delta_s/\Delta_o\approx 4$
between the spin  and the odd-mode gaps.
For this large ratio and in the regime $2 \Delta_o \ll T \cmin 2
\Delta_s$,
$\rho(T)$ is dominated  
 by the factor 
\mbox{$\propto T^{(\kr-3)/2}$}, 
since in this regime the fields $\bosp{\sp,\pm}$ have a vanishing
expectation value  \pcite{how}. 
This induces the down turn of $\rho(T)$  at these higher 
temperatures.

The  results shown by the full line in Fig. 2 may
provide
one possible explanation 
of the experimentally observed  maximum of 
$\rho(T)$ at $T\approx 100$ K in the normal state of 
\srca  \pcite{ue.na.96,params}. 
As mentioned above the large ratio $\Delta_s/\Delta_o$ guarantees the
experimentally observed down turn of $\rho(T)$ above its
maximum. Furthermore, our choice
of $\Delta_s/\Delta_o\approx 4$ is consistent with the following estimate:
from the full line in Fig.\ 2, $\Delta_s$ has a value of about $1.3$ times
the temperature for which $\rho(T)$ has a maximum. Comparing with
experiment, this fixes $\Delta_s$, i.e.\ $\Delta_s \approx 130$ K.
On the other hand, if we take 
the value for the undoped
 material, $\Delta_s(n=1)\approx 430$ K \cite{ue.na.96}, and estimate the
doping-induced reduction (according to the second of Ref.\
\cite{no.wh+scal}), we obtain $\Delta_s(n=0.75) \approx 0.33 \Delta_s(n=1)
\approx 140$ K. This is consistent with our previous result.
The  increase of $\rho(T)$ 
at $T\cmag 200$ K in the experiments
is probably due to contributions from electron-electron and electron-phonon
scatterings.
Notice that an appropriate description of 
the {\it superconducting} transition cannot be obtained by our simplified
 effectively one-dimensional model.
We finally comment on the pressure dependence of $\rho(T)$
shown
by the experiments. 
Application of hydrostatic pressure increases the carrier density in the
ladders\pcite{mo.os.97},  thus decreasing the overall resistivity (as
observed experimentally), and at the same time {\it slightly} reducing 
the values of the gaps and,
consequently, the   peak position in $\rho(T)$.

\ifpreprintsty\else\vspace*{-.3cm}\fi
\figb
\ifpreprintsty\else\vspace*{-.3cm}\fi

In conclusion, we have studied the transport properties of  
two-chain 
ladder systems in the presence of a barrier or of a low
impurity concentration.  We find a small region 
of repulsive interaction $0<U<U_c$
[cf. Fig. \ref{fig1}], for which the system is a perfect conductor 
for $T_{dis}\cmin T\cmin \Delta_o$.
We 
also suggested a possible explanation for the normal state behavior of
the resistivity in \srcax.

We thank J. Voit, E. Orignac, 
 H. J. Schulz, and R. M. Noack for stimulating discussions.
This work was partially supported by the EC network program
ERBCHRX-CT940438, by the Bavarian high-Tc program FORSUPRA 
and by the 
BMBF (05 605 WWA 6).
E. A. 
 acknowledges research support from the EC-TMR
  program  ERBFMBICT950048.

\ifx\undefined\andword \def\andword{and} \fi
\ifx\undefined\submitted \def\submitted{submitted} \fi
\ifx\undefined\inpress \def\inpress{in press} \fi
\def\nonformale#1{#1}
\def\formale#1{}
\def\spa{\ifpreprintsty\else\vspace*{-.5cm}\fi}
  \def\spb{\ifpreprintsty\else\vspace*{-1.6cm}\fi}
\spa

\endfigures


\begin{thebibliography}{10}
\spb

\bibitem{hald.81}{
F.~D.~M. Haldane, J. Phys. C {\bf 14},  2585  (1981).}%

\bibitem{ka+fu}{
\lcit{ka.fi.92}; \lcit{fu.na.93}.}%

\bibitem{ri.go.93}{
T.~M. Rice, S. Gopalan, and M. Sigrist, Europhys. Lett. {\bf 23},  445
  (1993).}%

\bibitem{no.wh+scal}{
\lcit{no.wh.95}; {\it ibidem}, cond-mat 9601047 (unpublished);
  \lcit{scal.nature}.}%

\bibitem{twoc}{
\lcit{fabr.93}; \lcit{kh.ri.94}; \lcit{schu.96}.}%

\bibitem{ba.fi.96}{
L. Balents and M.~P.~A. Fisher, Phys. Rev. B {\bf 53},  12133  (1996).}%

\bibitem{ec.ba+ko.ta}{
\lcit{ec.ba.94}; \lcit{ko.ta.95}.}%

\bibitem{ue.na.96}{
M. Uehara {\it et~al.}, J. Phys. Soc. Japan {\bf 65},  2764  (1996).}%

\bibitem{timp.90}{
G. Timp,  in {\em Mesoscopic Phenomena in Solids}, edited by B.~L. Altshuler,
  P.~A. Lee, and R.~A. Webb (Elsevier, Amsterdam, 1990).}%

\bibitem{notation}{
We are using here the notation of the last of Refs.\onlinecite{twoc}.}%

\bibitem{3c.pla}{
E. Arrigoni, Phys. Lett. A {\bf 215},  91  (1996).}%

\bibitem{unpub}{
Details will be given elsewhere: E. Arrigoni and B. Brendel and W. Hanke
  (unpublished).}%

\bibitem{or.gi}{
\lcit{or.gi.96}; {\it ibidem}, cond-mat 9704064.}%

\bibitem{c1s0}{
We restrict our analysis to values of the doping corresponding to the inner
  C1S0 phase of Ref.\onlinecite{ba.fi.96}.}%

\bibitem{ki.ao.95}{
T. Kimura, K. Kuroki, and H. Aoki, Phys. Rev. B {\bf 51},  13860  (1995).}%

\bibitem{v0v2}{
Here and in the following we deal, for simplicity, with the potentials $v_0$
  and $v_2$ only, since including the other terms (interband $v_{0,\pi}$ and
  other terms as relevant as $v_2$) does not change the form of $\rho(T)$ in
  Fig. 2.}%

\bibitem{uc}{
In particular, we estimate the contribution from (ii) by stopping the RG flow
  at the point $l=l^*$ where the strong-coupling regime is reached. To estimate
  the $U$-dependence of $\delta K_{\ch,+}$ we further add the {\it bare} values
  of the coupling \pcite{unpub}. This will be appropriate for $U\cmin \pi
  v_F$.}%

\bibitem{ba.fi.cn.96}{
L. Balents and M.~P.~A. Fisher, Phys. Rev. B {\bf 55},  R11973  (1997).}%

\bibitem{ya.sh.95}{
T. Yanagisawa, Y. Shimoi, and K. Yamaji, Phys. Rev. B {\bf 52},  R3860
  (1995).}%

\bibitem{tdis}{
All our results are valid down to $T\sim\tdis$.}%

\bibitem{ident}{
Unless otherwise specified, we consider $\Delta_o\sim \Delta_s$. Moreover, we
  assume both spin gaps to be equal.}%

\bibitem{valid}{
Assuming that the effective correlation exponents of the two spin and of the
  odd-charge modes are unity for $T\gg \Delta_s, \Delta_o$ and neglecting terms
  of order $(U/t \times \Delta_o v_F/v_F)$.}%

\bibitem{c2c0}{
The ratio $c_2/c_0$ can be calculated similarly to Ref. \onlinecite{ka+fu}. For
  the parameters of Fig. 2 we obtain $c_2/c_0\approx 2.61$.}%

\bibitem{params}{
Theoretical estimates place the \srcax compound into the strong-coupling region
  (M. Imada, S. Maekawa, {\it private communication}).}%

\bibitem{os.mo.96}{
T. Osafune {\it et~al.}, Phys. Rev. Lett. {\bf 78},  1980  (1997).}%

\bibitem{how}{
To evaluate $\rho(T)$ in the range $\Delta_o \le T \le \Delta_s$, we used the
  interpolation formula $v_0(T) \propto (T/E_F)^{(\kr+1)/4-1}
  (T+\Delta_s)^{1/2} \exp{-\Delta_o/T}$, which gives the correct behavior for
  $v_0(T)$ (and thus for $\rho(T)$) in the regions $T \gg 2 \Delta_s$, $
  2\Delta_o \ll T \ll 2\Delta_s$, and $T\ll 2 \Delta_o$ (see text).}%

\bibitem{mo.os.97}{
N. Motoyama {\it et~al.}, Phys. Rev. B {\bf 55},  R3386  (1997).}%

\end{thebibliography}
\end{document}